\documentclass[11pt]{article}
\pdfoutput=1

\usepackage{jcappub}
\usepackage{epsfig}

\usepackage{amsthm,graphicx,multirow}
\usepackage{epsfig}
\usepackage{latexsym, amssymb} 
\usepackage{amsmath}
\usepackage{epstopdf}
\usepackage[perpage]{footmisc}
\usepackage{soul}
\usepackage{hyperref}
\usepackage[dvipsnames]{xcolor}

\newcommand{\dd}{\mathrm{d}}

\begin{document}

\title{Testing Primordial Black Holes with multi-band observations of the stochastic gravitational wave background}

\author[a,b]{Matteo Braglia,}
\author[a]{Juan Garc\'ia-Bellido,}
\author[a,c]{Sachiko Kuroyanagi}

\affiliation[a]{Instituto de Fisica Teorica, Universidad Autonoma de Madrid, Madrid, 28049, Spain}
\affiliation[b]{INAF/OAS Bologna, via Piero Gobetti 101, I-40129 Bologna, Italy}
\affiliation[c]{Department of Physics and Astrophysics, Nagoya University, Nagoya, 464-8602, Japan}

\emailAdd{matteo.braglia@csic.es}
\emailAdd{juan.garciabellido@uam.es}
\emailAdd{sachiko.kuroyanagi@csic.es}

\abstract{ The mass distribution of Primordial Black Holes (PBHs) is affected by drops in the pressure of the early Universe plasma. For example, events in the standard model of particle physics, such as the $W^\pm/Z^0$ decoupling, the quark-hadron transition, the muon and pion becoming non-relativistic, and the annihilation of electrons and positrons, cause a suppression in the Equation of State parameter and leave peaks in the PBH mass function around $10^{-6},\,2,\,60$, and $10^6\, M_\odot$, respectively, in the case of a nearly scale-invariant primordial power spectrum. The superposition of unresolved mergers of such PBHs results in a stochastic gravitational-wave background (SGWB) that covers a wide range of frequencies and can be tested with future gravitational wave (GW) detectors. In this paper, we discuss how its spectral shape can be used to infer properties about inflation, the thermal history of the Universe, and the dynamics of binary formation in dense halos encoded in their merger rate formula. Although many of these physical effects are degenerate within the sensitivity of a single detector, they can be disentangled by the simultaneous observation of the SGWB at different frequencies, highlighting the importance of multi-frequency observations of GWs to characterize the physics of PBHs from the early to the late time Universe. }

\maketitle
      
\section{Introduction}

The development of Gravitational Wave (GW) astronomy has gone hand in hand with a thriving interest in Primordial Black Holes (PBHs). The first observation of a binary black hole (BBH) merger~\cite{LIGOScientific:2016sjg} has been shown compatible with scenarios in which PBHs constitute an appreciable fraction or even all of the cold dark matter (CDM)~\cite{Bird:2016dcv,Clesse:2016vqa,Sasaki:2016jop} and subsequent GW observations has uncovered tantalizing mysteries that may hint at the existence of PBHs. For example, the existence of compact objects in the so-called mass gap and the observed low effective spin of the mergers are features that can be hardly explained if black holes are of stellar origin, but are in perfect agreement with (and possibly predicted by) the PBH scenario~\cite{Clesse:2020ghq,DeLuca:2020sae,Garcia-Bellido:2020pwq}. Some scientists even believe the recent common-spectrum process observed by NANOGrav~\cite{NANOGrav:2020bcs} could also be a sign of PBHs~\cite{Vaskonen:2020lbd,DeLuca:2020agl,Kohri:2020qqd,Sugiyama:2020roc}.

Interestingly, GW astronomy is not the only channel to probe PBHs. They can be tested through their direct or indirect imprints on many more astrophysical and cosmological observables. The wealth of physical processes showing traces of their influence has been used to place tight bounds on the abundance of PBHs, which is becoming increasingly constrained~\cite{Carr:2020gox,Carr:2020xqk}. Such bounds, however, crucially depend on the mass function of the specific PBH model, which is often assumed to be monochromatic to simplify computations. In fact, broad mass functions are required to explain GW observations, which often relaxes other constraints, and PBHs can still constitute a sizeable fraction of the CDM.

On the theoretical side, many inflationary models can be responsible of the formation of PBHs with broad mass function (see e.g. Refs.~\cite{Dolgov:1992pu,Garcia-Bellido:1996mdl,Clesse:2015wea,Garcia-Bellido:2016dkw,Garcia-Bellido:2017mdw,Garcia-Bellido:2017aan,Palma:2020ejf,Fumagalli:2020adf,Braglia:2020eai,Braglia:2020taf,Ketov:2021fww,Khlopov:2008qy,Belotsky:2014kca,Kusenko:2020pcg,Kawai:2021edk}
 for an incomplete list). The usual approach is to produce a large peak in the primordial power spectrum of curvature fluctuations~\cite{Garcia-Bellido:1996mdl}. When the overdensities re-enter the Hubble radius, they collapse into PBHs if they are larger than a threshold $\delta_c$ and the resulting mass function qualitatively inherits the width and skewness of the peak in the primordial power spectrum. It is clear that, although physically viable, this approach requires some degree of fine-tuning in scale and amplitude of the peak if it is to explain observations. 

In this respect, another approach, less amenable to fine-tuning criticisms, has recently been explored~\cite{Carr:2019hud,Garcia-Bellido:2019vlf,Carr:2019kxo}. Considering changes in the pressure of the early Universe plasma, enhanced gravitational collapse occurs because $\delta_c$ decreases with the slight softening of the equation of state (EOS) parameter $w$. Therefore, this approach does not require the presence of a large peak, but simply a featureless power spectrum with an amplitude that has not to be so large\footnote{This strictly holds for Gaussian fluctuations. For the effects of non-Gaussianities, see e.g. Refs.~\cite{Young:2013oia,Garcia-Bellido:2017aan,Franciolini:2018vbk,Atal:2018neu,DeLuca:2019qsy,Yoo:2019pma,Ezquiaga:2019ftu} .}. A very appealing aspect of this model is that it might explain the nature of CDM and the LIGO-Virgo-Kagra (LVK) results with only elements of the SM of particle physics. For example, the Higgs field playing the role of the inflaton can produce a (near) plateau in the primordial power spectrum by slowing down after crossing a near-inflection point related to the critical value of the running of its self-coupling and non-minimal coupling to gravity~\cite{Ezquiaga:2017fvi}. Then, SM particles becoming non-relativistic produce sudden drops in the EOS that cause the mass function to show peaks around $10^{-6},\,2,\,60$ and $10^6\, M_\odot$. Such peaks are a key prediction of the model.

In this paper, we show that this scenario also leaves some very distinct imprints on the Stochastic Gravitational-Wave Background (SGWB). Unlike other models, the very broad mass function leads to the formation of primordial BBHs with total masses and mass ratios across many orders of magnitudes. The SGWB resulting from such mergers in this scenario drastically deviates from the $\Omega_{\rm GW}\sim f^{2/3}$ behavior, typical of astrophysical black holes, and its amplitude across several decades in frequency is large enough to fall in the sensitivity range of many planned GW detectors.

Our purpose is to carefully study the SGWB from primordial BBH mergers and understand what physical information can be extracted from its spectral shape. We demonstrate that the SGWB is a powerful probe of many physical effects occurring from primordial to late times. The SGWB produced in this model can therefore be used not only to test the clustering properties of PBHs, but also the inflationary scenario that produced the primordial overdensities, as well as the possibility that exotic phase transitions happened around the QCD quark-hadron transition.  Crucial to our analyses is the multiwavelength approach. Indeed, although different physical mechanisms imprint degenerate features within the frequency range of many future detectors, such a degeneracy can be broken partially or completely by observing the SGWB at different frequencies.

We stress that, although the SGWB from PBH binaries has been studied in many papers (see e.g. Refs.~\cite{Clesse:2016ajp,Raidal:2017mfl,Raidal:2018bbj,Wang:2019kaf,Mukherjee:2021ags,Mukherjee:2021itf,Zagorac:2019ekv,Franciolini:2021tla,Bavera:2021wmw}), this is the first case where it is computed by taking into account the effects of the thermal history of the Universe on the PBHs mass function. 

The paper is structured as follows. We begin by reviewing the model and constructing a phenomenological expression for the merger rate of primordial BBHs in Section~\ref{sec:model}. In the following Section~\ref{sec:SGWB}, we compute the SGWB, carefully analyzing the imprint of the primordial spectrum, the dynamics of binary formation, and the thermal history of the Universe across the scales probed by ground and space-based GW interferometers as well as Pulsar Timing Arrays (PTA) and high precision astrometry (GAIA/THEIA). Finally, in Section~\ref{sec:pls}, we use the binned power-law sensitivity (PLS) curve formalism to qualitatively demonstrate the power of such experiments to reconstruct the spectral shape of the SGWB. We draw our conclusions in Section~\ref{sec:conclusion}.

\section{Model and merger rate}
\label{sec:model}

\begin{figure}
	\begin{center}
		\includegraphics[width=.495\columnwidth]{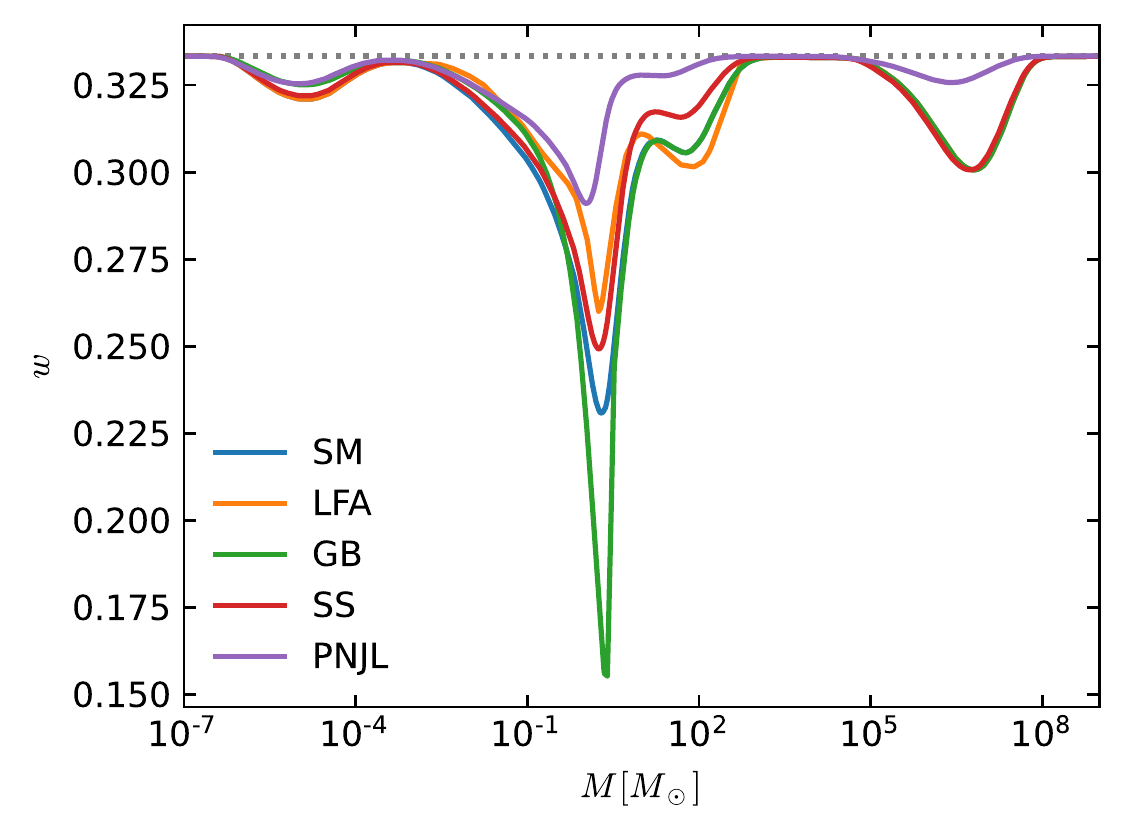}
		\includegraphics[width=.495\columnwidth]{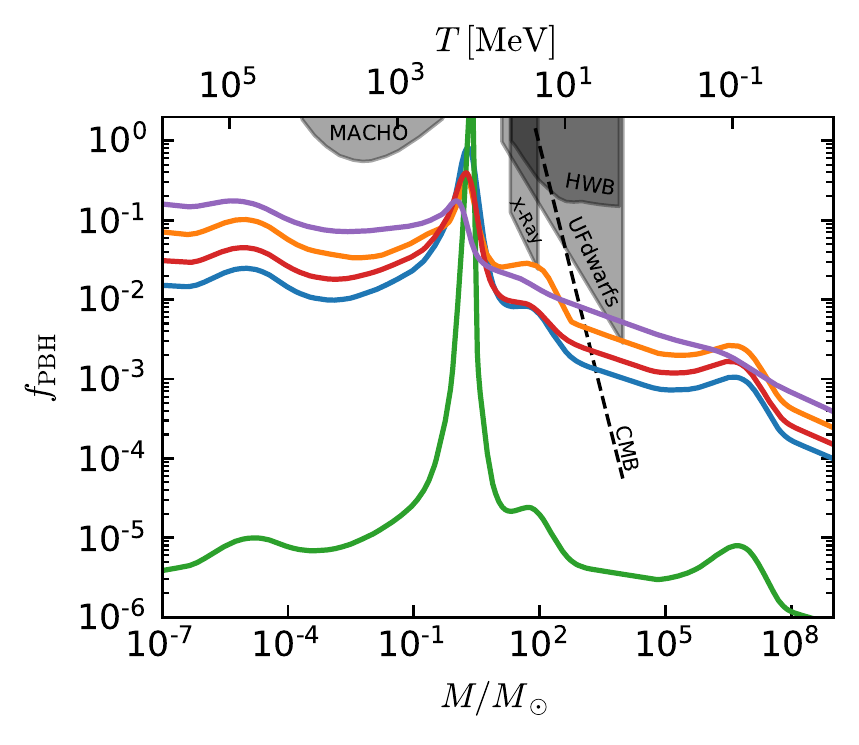}
	\end{center}
	\caption{\label{fig:fpbh}[Left] EOS parameter $w$ as a function of the temperature $T$ for different models of the thermal history of the Universe. The standard value $w=1/3$ in the radiation-dominated era is plotted with a gray dotted line. [Right] The mass spectrum of PBHs for different thermal history models. To produce the plot, we normalize the mass function with $f^{\rm TOT}_{\rm PBH}=1$, and assume a spectral index $n_s=0.97$. Shaded regions are constraints from microlensing (MACHO), ultra-faint dwarf galaxies, and Eridanus II (UDFdwarfs)~\cite{DES:2016vji}, X-ray/radio counts (X-Ray)~\cite{Gaggero:2016dpq}, and halo wide binaries (HWB)~\cite{Quinn:2009zg}. The accretion constraint (CMB)~\cite{Ali-Haimoud:2016mbv,Poulin:2017bwe,Serpico:2020ehh} is shown with the dashed line because the modeling of accretion adopted to derive it breaks down at the large masses ($\sim10^4\,M_\odot$). The upper $x$-axis shows the temperature scale associated with the PBH mass. Since $T(M)$ depends on the number of relativistic degrees of freedom, which depends on the model, we use the approximate relation $T\sim200\sqrt{M_\odot/M}$ MeV in the plot. }
\end{figure}

The formation of PBHs requires the generation of large overdensities in the early Universe, which are usually taken to be of inflationary origin. When such overdensities at horizon re-entry are larger than the threshold $\delta_{c}\equiv \delta \rho / \rho$, they collapse into PBHs. The specific value of the critical density is well known to be very sensitive to the profile of the EOS of the cosmic fluid at the time when the perturbations enter the Hubble radius. Therefore, different models of the thermal history evolution can produce distinct features in the PBH mass function through their imprints on $w(T)$. Indeed, the fraction of horizon patches that collapse into PBHs is given by~\cite{Carr:1975qj}
\begin{equation}
\beta( M )
\approx
{\rm erfc}\!
\left[
\frac{\delta_{c}\big( w[ T( M ) ] \big)}
{ \sqrt{2} \, \delta_{\rm rms}( M )}
\right]
,
\label{eq:beta(T)}
\end{equation}
where `erfc' is the complementary error function, $\delta_{\rm rms}( M )$ is the root-mean-square amplitude of the inhomogeneities at the mass scale $M$, and the temperature $T$ is related to the mass of PBHs by the relation
\begin{equation}
\label{eq:TofM}
T
\approx
\sqrt{\gamma}\,700\,g_*^{-1/4}
\sqrt{M_\odot / M\,}\;{\rm MeV}
\, ,
\end{equation}
where $g_*(T)$ is the number of relativistic degrees of freedom (which depends on temperature), and $\gamma$ parametrizes the ratio between the PBH mass and the mass of the collapsing horizon-sized region at PBH formation, which typically takes values $\gamma\in[0.2,\,1]$. In the following, we fix $\gamma=0.8$ and use the numerical results of Ref.~\cite{Musco:2012au} for $\delta_c(w)$.

Starting from Eq.~\eqref{eq:beta(T)}, we can compute the mass fraction of CDM in the form of PBHs per logarithmic interval of masses using the following formula:
\begin{equation}
f_{\rm PBH}( M ) 
\approx 2.4\,\beta( M ) \sqrt{\frac{M_{\rm eq}}{M}}
\, ,
\label{eq:fPBH}
\end{equation}
where $M_{\rm eq} = 2.8 \times 10^{17}\,M_\odot$ is the horizon mass at matter-radiation equality. The numerical factor is $2( 1 + \Omega_b / \Omega_{\rm CDM} )$, with $\Omega_{\rm CDM} = 0.245$ and $\Omega_b = 0.0456$ being the CDM and baryon density parameters~\cite{Planck:2018vyg}. We will often call $f_{\rm PBH}( M )$ \emph{mass function} in the following. By integrating $f_{\rm PBH}( M )$ over its full support we get the total fraction of PBH:
\begin{equation}
\label{eq:ftot}
f_{\rm PBH}^{\rm tot}\equiv\int\,d\ln M \,f_{\rm PBH}( M ).
\end{equation}

As stated in the Introduction, we assume Gaussianity of the perturbations with root-mean-square $\delta_{\rm rms}( M )$, which we parameterize as follows
\begin{equation}
\delta_{\rm rms}( M )
=
\begin{cases}
\tilde{A}
\left(
\frac{ M }{ M_\odot }
\right)^{\!(1 - \tilde{n}_{s}) / 4}\, 10^{-\frac{\tilde{\alpha}_s}{2}\log_{10}^2 \frac{M}{M_\odot}}
&\; \text{(PBH scales)} \\[2mm]
A
\left(
\frac{ M }{ M_\odot }
\right)^{\!(1 - n_{s}) / 4}
&\; \text{(CMB scales).} 
\end{cases}
\label{eq:delta-power-law}
\end{equation}
The spectral index $n_s$ and amplitude $A$ at CMB scales should be consistent with the CMB observations~\cite{Planck:2018vyg}, but we have more freedom to change the perturbations at small (PBH) scales. 
As can be read from this equation, we allow variation of both the tilt $\tilde{n}_s$ of the power spectrum at PBH scales and its running $\tilde{\alpha}_s$. Models such as critical Higgs inflation can indeed be better represented by a non-vanishing running of the spectral index~\cite{Ezquiaga:2017fvi,Hasinger:2020ptw}. In the remaining part of this Section, we will take the fiducial values to be $\tilde{n}_s=0.97$ and $\tilde{\alpha}_s=0$, and analyze their variation in the following Section. The amplitude $\tilde{A}$ is determined by fixing $f_{\rm PBH}^{\rm tot}=1$ using  Eq.~\eqref{eq:ftot}. 

In this paper, we will consider several variations of the thermal history of the Universe and  evolution of the EOS parameter $w(T)$ shown 
in the left panel of Fig.~\ref{fig:fpbh}. Apart from the thermal evolution of the Standard Model (SM) of particle physics, we consider also the following beyond SM models
	 Lepton Flavor Asymmetries (LFA) of Ref.~\cite{Bodeker:2020stj} (their case ({\em iii}) ) and the 
	 Glueballs (GB),  Solitosynthesis (SS) and  PNJL models, for which we explore the benchmark cases of Fig.~6
of Ref.~\cite{Garcia-Bellido:2021zgu} (for the GB model we consider $N_C=5$ colors).

With our choice of $\gamma=0.8$, we observe a prominent peak around $M\sim 2 M_\odot$ because of the quark-hadron transition. A different value of $\gamma$ would slightly shift the peak up and down in mass. The mass function of the SM shows three secondary bumps at  $\sim10^{-6}M_\odot,\,\sim60 M_\odot$ and $\sim10^{6} M_\odot$.  All these peaks arise because of the sudden drops in the effective EOS $w$ in the early Universe.
Since the process of PBH formation is exponentially dependent on the  density threshold $\delta_c$, which in turn is a function of $w$, we see a peak in correspondence of PBHs forming when $w$ drops. Relating the temperature of the Universe to the typical PBH mass by Eq.~\eqref{eq:TofM}, the bumps in the mass function can be easily understood (see also the scale of the top x-axis). Going from smaller to larger masses, the first bump is due to  the top quark, Higgs boson and Z and W bosons becoming non-relativistic, whereas the main bump at $M\sim 2 M_\odot$ corresponds to the QCD transition and is followed by a small bump for PBHs forming when pions become non relativistic, and a fourth bump for PBHs with $10^6 M_\odot$ that form when positrons and electrons annihilate.

The impact of different thermal histories on the mass function can be quite dramatic, as can be seen from Fig.~\ref{fig:fpbh}, and mainly consists in changing the relative importance of the four peaks. Notable cases are the PNJL model, where the main peak is almost washed out, and the GB model, where instead a very sharp drop in $w$ around the quark-hadron transition enhances the main peak at $M\sim2 M_\odot$ which is  $5$ orders of magnitude above the others.

\subsection{Merger rate}

In order to compute the SGWB produced by the mergers of unresolved binaries, we need an expression for their merger rate. Our derivation relies on the assumption that PBH binaries formed by tidal capture in very dense halos during the matter dominated era. However, we note that other scenarios exist where binaries can form in the early Universe~\cite{Sasaki:2016jop,Clesse:2020ghq}\footnote{For the effects of different assumptions on the mechanism of binary formation on the SGWB, see Ref.~\cite{Bagui:2021dqi}, which appeared on the arXiv simultaneously to this paper. We note that, although their overall findings qualitatively agree with ours, the $\Omega_{\rm GW}$ they compute is quantitatively different due to the use of a slightly different formula for the computation of the SGWB, which does not include the merger and ringdown phases.}

Within the scenario just discussed, we model the merger rate as~\cite{Clesse:2016vqa}:
\begin{equation}
\label{eq:tauz}
\frac{\dd^2\tau_{\rm merg}(z,\,m_1,\,m_2)}{\dd\log_{10} m_1\,\dd\log_{10} m_2}=\overbrace{R_{\rm clust}\frac{(m_1 + m_2)^{10/7} }{(m_1 m_2)^{5/7}}\,\frac{(1+z)^{\alpha_z}}{\left[1+\left(M_{\rm tot}/M_*\right)\right]^{\alpha_c}}}^{\rm PBH\,clustering/binary\, formation}\,\overbrace{f_{\rm PBH}(m_1)f_{\rm PBH}(m_2)}^{\rm primordial / thermal\, evolution}.
\end{equation}
As shown explicitly in the equation, not only does the merger rate depend on the properties of PBH clustering and their binary formation, but it also depends on the early Universe physics through the factor $f_{\rm PBH}(m_1)f_{\rm PBH}(m_2)$, which we have already discussed in the previous section.

Let us now discuss our merger rate formula following Refs.~\cite{Mukherjee:2021ags,Mukherjee:2021itf}., we insert a factor $(1+z)^{\alpha_z}$, where $\alpha_z$ is a positive index, as an Ansatz to parametrize our ignorance about clustering dynamics.  Typical values of the index $\alpha_z$ include $\alpha_z=0$, i.e. a constant merger rate, as assumed in Refs.~\cite{Clesse:2016vqa,Clesse:2016ajp,Clesse:2020ghq} or $\alpha_z=1.3$ as in Refs.~\cite{Raidal:2017mfl,Raidal:2018bbj}

$R_{\rm clust}$ is a constant parameter of dimension ${\rm yr}^{-1} {\rm Gpc}^{-3}$ which encodes the clustering properties of PBHs. In the following, we consider $R_{\rm clust}$ as a free parameter in our model and, unless stated otherwise, we fix it by requiring that the integral of the merger rate at $z=0$ over the mass range $[5,\,100]\,M_\odot$ gives a total rate of $\sim 38\,{\rm yr}^{-1}\,{\rm Gpc}^{-3}$, 
consistent with the edge of the 1$\sigma$ constraint on the local merger rate inferred from GWTC-2~\cite{LIGOScientific:2020kqk}. We note that the parameter governing the amplitude of the merger rate is actually $R_{\rm clust}\,(f_{\rm PBH}^{\rm tot})^2$ and $R_{\rm clust}$ is thus degenerate with the the total fraction of PBHs, as also noted in Refs.~\cite{Raidal:2017mfl,Vaskonen:2019jpv,Young:2019gfc,Trashorras:2020mwn,Atal:2020igj,DeLuca:2020jug}. However, since we normalize $f_{\rm PBH}^{\rm tot}=1$, we break this degeneracy by hand.

\begin{figure}
	\begin{center}
		\includegraphics[width=.7\columnwidth]{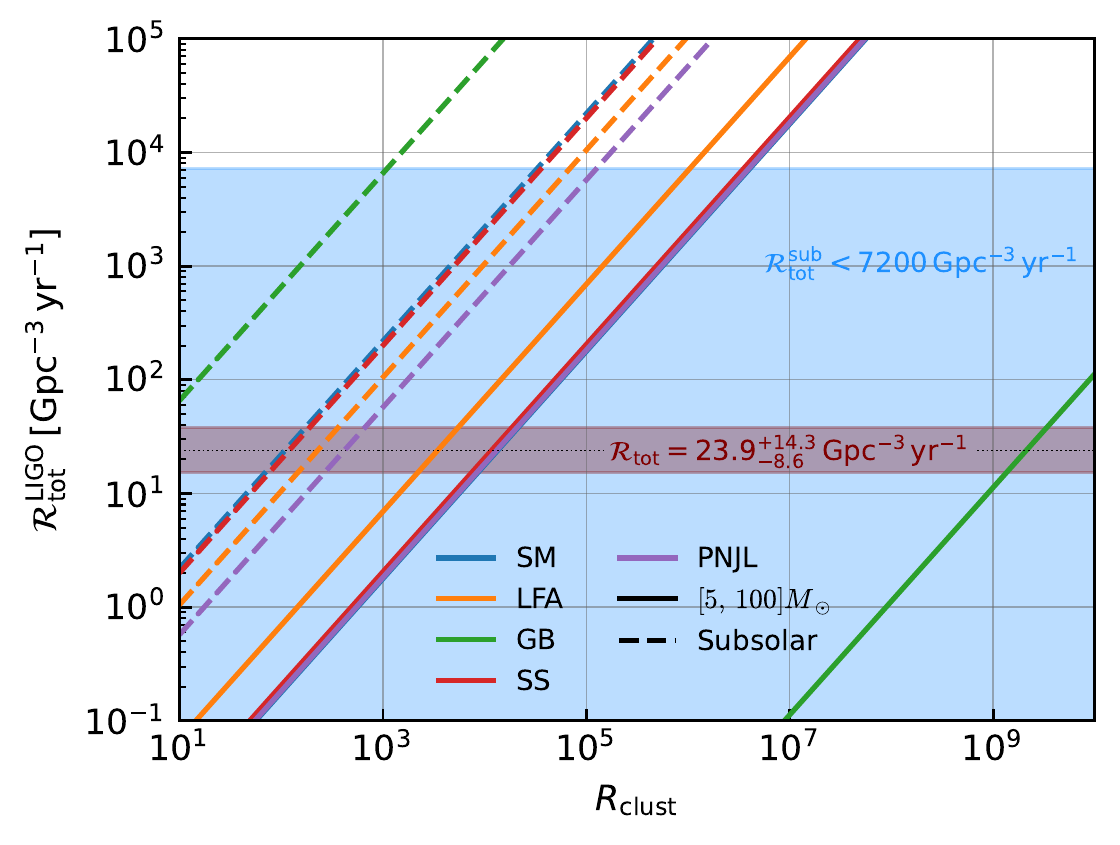}
	\end{center}
	\caption{\label{fig:tau}  Dependence of the merger rate at the LVK frequency band on different thermal history models as a function of $R_{\rm clust}$. The blue shaded region corresponds to the upper bound on the merger rate of subsolar mass BBHs obtained in~\cite{Nitz:2021vqh}, while the brown shaded region is the merger rate inferred by the GWTC-2 catalog~\cite{LIGOScientific:2020kqk}. Dashed lines show the total rate for subsolar mass PBHs, whereas solid ones show the total rate of PBHs in the mass range of $[5,\, 100]\, M_\odot$.}
\end{figure}

In Fig.~\ref{fig:tau}, we show how the merger rate at the LVK sensitivity changes depending on different thermal histories as a function of $R_{\rm clust}$. We plot the merger rate of the mass range $[5,\,100]\,M_\odot$ corresponding to the search range of the GWTC-2 catalog (solid lines) and $[0.1,\,7]\,M_\odot$ corresponding to the search range of subsolar mass BBHs~\cite{Nitz:2021vqh,LIGOScientific:2021job} (dashed lines). The behavior of solid lines is very similar for all models except for the LFA and the GB models, whose mass function significantly differs from the SM model in the range $[5,\,100]\,M_\odot$. We see that the pion peak at $\sim 80 M_\odot$ is broader in the LFA case and less pronounced in the GB one, resulting in a smaller and larger value of $R_{\rm clust}$ to match the LVK rate,  respectively.
 
In Table~\ref{tab:RclustSub}, we list the values of $R_{\rm clust}$ satisfying the upper bounds of the merger rate indicated by GWTC-2 and subsolar mass search. For subsolar mass binaries, following~\cite{Nitz:2021vqh}, we normalize the merger rate integrating the primary mass over $[0.1,\,7]\,M_\odot$ and the secondary one over $[0.1,\,1]\,M_\odot$, and we compute $R_{\rm clust}$ by requiring the total merger rate to be equal to $7200\,{\rm yr}^{-1}\,{\rm Gpc}^{-3}$. For all the models, the values of $R_{\rm clust}$ obtained using the two normalizations are very similar, leading to the conclusion that the PBH model can explain both rates. However, for the GB model, the two are in tension. Since subsolar mass BHs, if they exist, are very likely of primordial origin, we choose to normalize the GB model to the subsolar mass rate in the following. This means that this specific model cannot explain the LVK rate of BBHs of large masses, which is therefore attributed to astrophysical black holes.

\begin{table}
	\centering
	\begin{tabular}{|l|l|l|}
	\hline
	Model               & Subsolar&   $[5,\,100]\,M_\odot$    \\ \hline
	SM   & $3.3\times 10^4$ &$2.1\times 10^4$    \\ \hline
	LFA   & $6.8\times 10^4$ &  $5.5\times 10^3$   \\ \hline
	GB   & 1100& $3.3\times 10^9$  \\ \hline
	SS   & $3.6\times 10^4$ &$1.8\times 10^4$    \\ \hline
	PNJL   &$1.2\times 10^5$   &$2.1\times 10^4$   \\ \hline
\end{tabular}
	\caption{\label{tab:RclustSub} Values of $R_{\rm clust}$ for different thermal history models, satisfying the upper bounds of the merger rate indicated by GWTC-2 and subsolar mass search. Here, we fix $f_{\rm pbh}^{\rm tot}=1$. In the first column, we obtain the value of $R_{\rm clust}$ by requiring consistency with the upper bound of the rate of subsolar-masses binaries, i.e. a total merger rate of $7200\,{\rm yr}^{-1}\,{\rm Gpc}^{-3}$ when integrating the primary mass between $[0.1,\,7]\,M_\odot$ and the secondary one between $[0.1,\,1]\,M_\odot$. In the second column, we compute $R_{\rm clust}$ by requiring that the integral of the merger rate at $z=0$ over the mass range $[5,\,100]\,M_\odot$ gives a total rate of $38\,{\rm yr}^{-1}\,{\rm Gpc}^{-3}$.}
\end{table}

Strictly speaking, the constancy of $R_{\rm clust}$ over the very broad mass range the PBH mass function in our scenario can be a poor approximation. In particular, we can expect the merger rate to decrease at large $M_{\rm tot}\equiv m_1+m_2>10^6\,M_\odot$, since the dynamics of clusters~\cite{Trashorras:2020mwn} make the more massive PBHs become isolated and possibly dressed by gas in accretion disks, thus rarely meeting each other until late-time galaxy mergers occur.
In order to model such a suppression, we impose a cutoff in the merger rate at large total masses by inserting the factor $1/\left[1+\left(M_{\rm tot}/M_*\right)\right]^{\alpha_c}$, where $\alpha_c$ is another positive index and we compare the cases of $M_*=10^5 M_\odot$ and $10^6 M_\odot$ in the following section.

\section{Stochastic Gravitational Wave Background from PBH mergers}
\label{sec:SGWB}

\begin{figure}
	\begin{center}
		\includegraphics[width=.75\columnwidth]{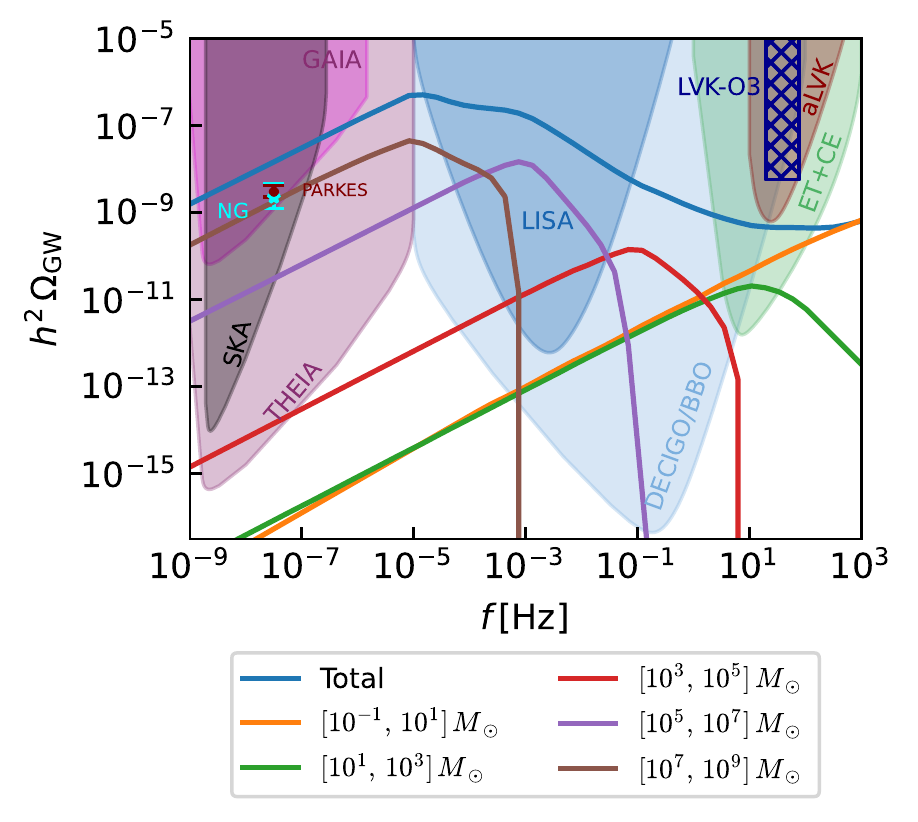}
			\end{center}
			\caption{\label{fig:omega} The SGWB generated in the SM scenario with the spectral index of $n_s=0.97$ and no running. We also show the contribution of different mass bins to $\Omega_\text{\rm GW}$. We also show the PLS curves with ${\rm SNR}=10$ for SKA, GAIA, THEIA, LISA, BBO, advLIGO+Virgo+KAGRA (aLVK) and ET+CE as well as the implications on the amplitude of the SGWB at $f=1 {\rm yr}^{-1}$ by NANOGrav and Parkes PTA and the upper bound from the O3 run of LVK in the range $f\in[20,\,76.6]$ Hz.  We assume the SM thermal history and a constant merger rate.}
\end{figure}

Starting from the merger rate, we follow the standard procedure to derive the SGWB formed by overlapped GWs of compact binaries, see e.g.~\cite{Rosado:2011kv}. The present day energy fraction of the SGWB is defined by $\Omega_\text{\tiny GW} \equiv (d \rho_{\rm GW}/ d\ln f)/\rho_c$, where $\rho_{\rm GW}$ is the energy density of GWs, and it is given by the following integral:
\begin{equation}
\label{eq:Omega_z}
\Omega_\text{\tiny GW} (f)=  \frac{f}{\rho_c} \int_0^{z_{\rm max}} \dd z'\, \dd m_1 \,\dd m_2 \,\frac{1}{(1+z')H(z')}\times  \frac{\dd^2 \tau_{\rm merg}(z',\,m_1,\,m_2)}{\dd m_1 \dd m_2}  \frac{\dd E_\text{\tiny GW} (f_s)}{\dd f_s},
\end{equation}
where $z_{\rm max}=\frac{f_3}{f}-1$ and $f_s$ is the redshifted source frame frequency $f_s =f (1+z)$, $\rho_c = 3 H_0^2/8\pi G$ is the critical energy density of the Universe. In the non-spinning limit, the single source energy spectrum is given by the following phenomenological expression~\cite{Ajith:2009bn} 
\begin{equation}
	\frac{\dd E_\text{\tiny GW} (f)}{\dd f} =  \frac{\pi^{2/3}}{3} (G\mathcal{M}_c)^{5/3} f^{-1/3} \times
	\left \{ \begin{array}{rl}
	& \left( 1+ \alpha_2 u^2\right)^2 \qquad \qquad \quad  \text{for} \quad f < f_1,  \\
	& w_1 f \left( 1+ \epsilon_1 u + \epsilon_2 u^2 \right)^2 \qquad  \text{for} \quad f_1 \leq f < f_2, \\
	& w_2 f^{7/3} \frac{ f_4^4}{(4 (f- f_2)^2 + f_4^2)^2} \quad \quad \quad \quad \ \ \text{for}  \quad f_2 \leq f < f_3 , 
	\end{array}
	\right.
\end{equation}
where $u_{(i)}\equiv(\pi M_{\rm tot} G f_{(i)} )^{1/3}$, $\alpha_2 = -{323}/{224} + \eta\, {451}/{168}$, $\eta = m_1 m_2/M_\text{\tiny tot}^2$, $\epsilon_1 = -1.8897$, $\epsilon_2 = 1.6557$,
\begin{align}
&w_1  = f_1^{-1} \frac{[1 + \alpha_2 u_1^2]^2}{[1+ \epsilon_1 u_1 +\epsilon_2 u_1^2]^2},  \nonumber \\
&w_2  = w_1 f_2^{-4/3} [1+ \epsilon_1 u_2 +\epsilon_2 u_2^2]^2,  
\end{align}
and
\begin{align}
&u_1^3  = 0.066+0.6437\eta-0.05822\eta^2-7.092\eta^3,  \nonumber \\
&u_2^3  = 0.37/2+0.1469\eta-0.0249\eta^2+2.325\eta^3,  \nonumber \\
& u_3^3 = 0.3236 -0.1331\eta -0.2714\eta^2 +4.922\eta^3,\nonumber\\
&u_4^3 = (1-0.63)/4 -0.4098\eta +1.829\eta^2-2.87\eta^3.
\end{align}

As an example, in Fig.~\ref{fig:omega}, we plot the SGWB from primordial BBHs produced in the context of the SM thermal history, with the primordial power spectrum of $\tilde{n}_s=0.97$ and no running. We compare it with the Power-Law-Sensitivity (PLS) curves~\cite{Thrane:2013oya} of future GW experiments obtained by requiring a signal-to-noise ratio (SNR) of the measured SGWB of ${\rm SNR}=10$ (see Section~\ref{sec:pls} for a detailed description of how the PLS is computed). As we can see in the figure, future GW experiments can cover the wide range of frequencies from $10^{-9}$ to $10^3$ Hz. At low frequencies, we consider SKA, which will work as a PTA observatory, and GAIA~\cite{Gaia:2018ydn} and THEIA, which will be used as GW observatories by monitoring the apparent position of stars (see e.g. Refs.~\cite{Book:2010pf,Moore:2017ity,Mihaylov:2018uqm,Garcia-Bellido:2021zgu}). At intermediate frequencies, we show the PLS curves of two space-based observatories, i.e. LISA and DECIGO/BBO.  Finally, at the higher frequencies, we show the PLS for combinations of advanced LVK and of the Einstein Telescope and the Cosmic Explorer (ET+CE), as considered in Ref.~\cite{Alonso:2020rar}.  We note, however, that several other GW experiments are planned that have sensitivity ranges overlapping with those considered here. In addition to the PLS curves, we also plot the constraints on the amplitude of the SGWB obtained by assuming that it is responsible for the stochastic common process signal detected by NANOGrav~\cite{NANOGrav:2020bcs} and the Parkes PTA~\cite{Goncharov:2021oub} and the upper bound from the LVK O3 run in the range of $[20,\,77]$ Hz~\cite{KAGRA:2021kbb}.

As can be seen, $\Omega_{\rm GW}$ extends over a very broad range of frequencies. Furthermore, the resulting spectral shape of $\Omega_{\rm GW}$ is quite peculiar and is very different from the one shown in the literature for other PBH models. Indeed, the typical shape of $\Omega_{\rm GW}$ consists in a $f^{2/3}$ scaling at small frequencies followed by a sharp cutoff at higher frequencies. As can be seen from the expression above, the scale of the cutoff is roughly determined by the frequency of the innermost stable circular orbit, which scales as the inverse of the total mass of the binary. Therefore, the power in $\Omega_{\rm GW}$ from binaries with larger $M_{\rm tot}$ drops at small frequencies and vice versa.

In order to better understand the origin of the spectral shape of the SGWB in our model, we have broken down the full support of the PBH mass function into smaller mass bins and computed the SGWB for each of them. The result, also shown in Fig.~\ref{fig:omega}, looks now more familiar and clearly explains the origin of the unusual spectral shape of the full $\Omega_{\rm GW}$, which is qualitatively the superposition of the results in each mass bin. We now go on and look in detail at how the variation of primordial physics, thermal history, and the clustering properties of PBHs modify such a SGWB.

We note that the SGWB computed here is not subjected to the bound from Big Bang Nucleosynthesis since most of the binaries merge after recombination, and thus the corresponding SGWB does not contribute to extra relativistic degrees of freedom during the early Universe. Furthermore, although the specific SGWB shown in Fig.~\ref{fig:omega} is in tension with the NANOGrav and Parkes PTA results, we stress that our focus here is to clearly show the effects of different physical mechanisms on the SGWB. In fact, as we will see in the next subsections, some values of the theory parameters bring the SGWB back into agreement with PTA results. 

\begin{figure}[t]
	\begin{center}
		\includegraphics[width=.495\columnwidth]{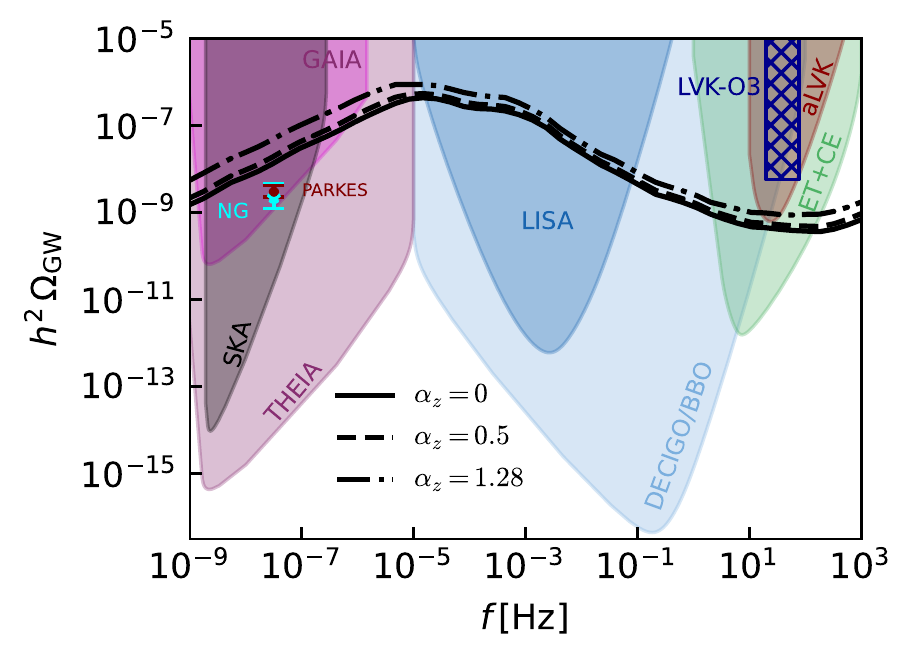}		\includegraphics[width=.495\columnwidth]{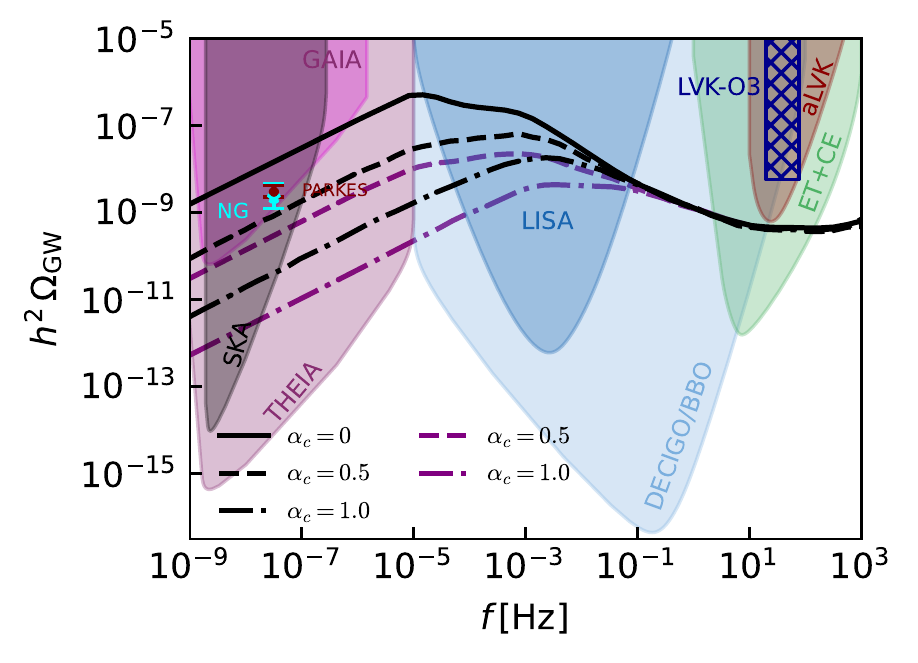}
	\end{center}
	\caption{\label{fig:astro} [Left] Effects of varying the redshift dependence $\alpha_z$ of the merger rate and [right] the cutoff exponent $\alpha_c$ on the SGWB. In the right panel, black and purple lines correspond to $M_*=10^6\,M_\odot$ and $M_*=10^5\,M_\odot$, respectively.}
\end{figure}

\subsection{The dynamics of binary formation}

We start by analyzing the imprints of different assumptions on the merger rate on the SGWB. The effects that we consider here are simply to change the redshift evolution of the merger rate and the small frequency cutoff. We show our results in Fig.~\ref{fig:astro}.

As can be seen, larger values of the exponent $\alpha_z$, governing the growth of the merger rate as we go back in redshift, are accompanied by a large amplitude of $\Omega_{\rm GW}$ (see also in Ref.~\cite{Mukherjee:2021ags}).

An increasing cutoff exponent $\alpha_c$ suppresses the power in the SGWB at LISA to PTA/THEIA frequencies. In the most extreme case,  the bump around $f\sim10^{-5}$~Hz is completely erased, and the SGWB in the LISA sensitivity band, which can be well described by a broken power-law, may be confused with signals coming from GW production in phase transitions (see e.g.~\cite{Caprini:2019egz}).

We note that suppression of power at very small frequencies is also expected if binaries have an initially high eccentricity~\cite{Clesse:2016ajp}.

\subsection{Primordial fluctuations from Inflation}

We now analyze the imprints of the primordial power spectrum of curvature perturbations on the SGWB. As already discussed, we only consider the effects of tilting or very slightly distorting the  scale-invariant power spectrum. More peaked power spectra, such as those described by a narrow lognormal distribution, would hardly show the effects of changes in the EOS, and our results would thus be very similar to those already obtained in the literature, see e.g.~\cite{Wang:2019kaf}.

\begin{figure}
	\begin{center}
		\includegraphics[width=.495\columnwidth]{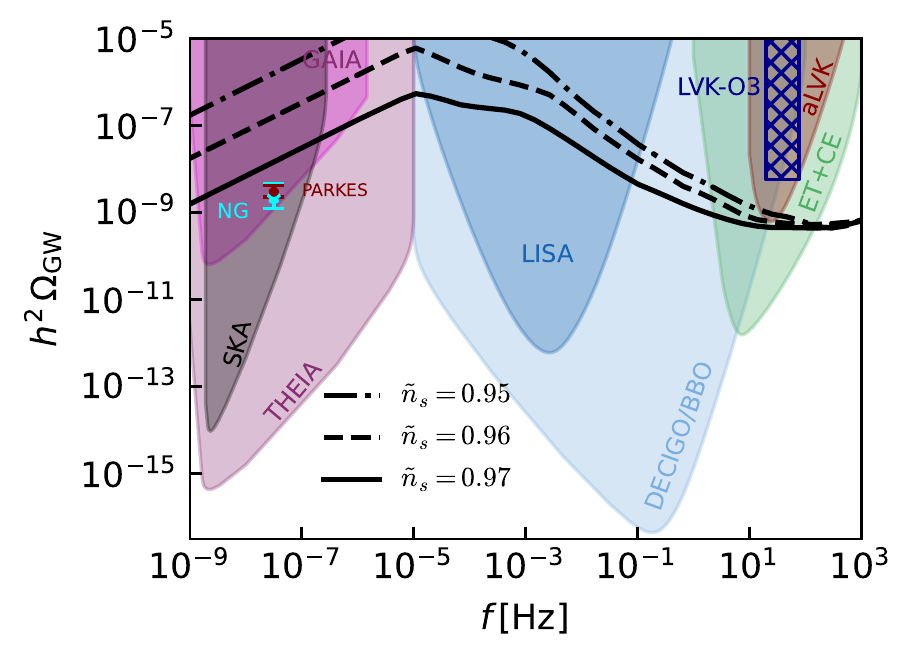}		\includegraphics[width=.495\columnwidth]{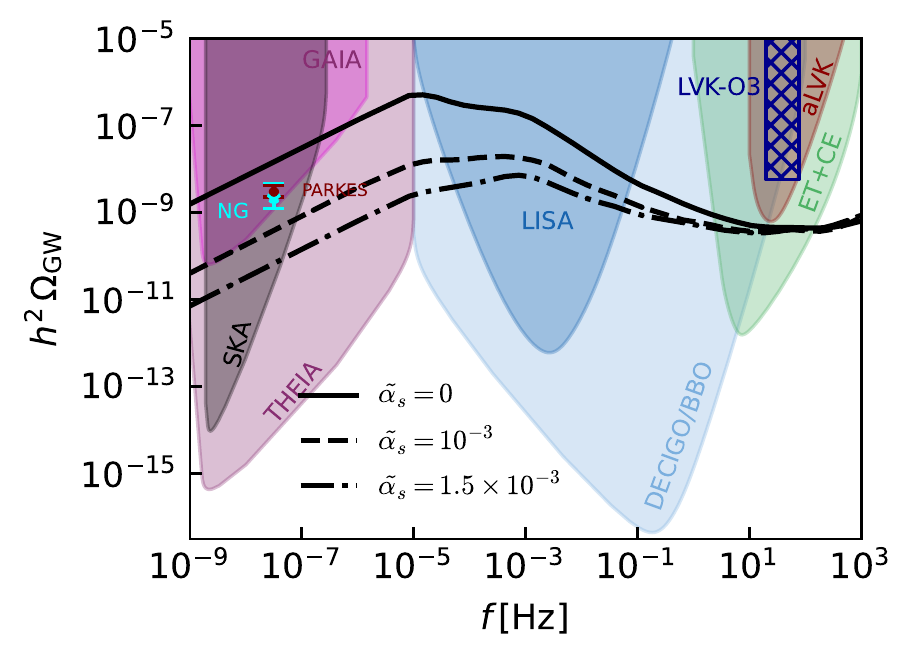}
	\end{center}
	\caption{\label{fig:tilt} [Left] Effects of varying the tilt $\tilde{n}_s$ of the primordial power spectrum and [right] its running $\tilde{\alpha}_s$ on the SGWB. We assume the SM thermal history and a constant merger rate with redshift.}
\end{figure}

Our results are shown in Fig.~\ref{fig:tilt}. At first sight, the effects of varying the tilt and the running of the primordial power spectrum look very similar. However, a more accurate inspection shows that adding some extent of running, the two peaks around $f\sim10^{-5}$ and $10^{-3}$ Hz disappear because they are smoothed out in the mass function. This highlights the importance of the multiwavelength analysis of the SGWB. Indeed, if we restricted to e.g. the sensitivity band of ground-based detectors, such as the combination of ET and CE, there would be an almost complete degeneracy between the effects of varying the tilt, and it would be impossible to extract any sensible information about primordial physics from the SGWB. Such a degeneracy, however, can be broken by reconstructing the spectral shape of the SGWB with space-based detectors and/or PTA experiments.

\subsection{Thermal History of the Universe}
\begin{figure}
	\begin{center}
		\includegraphics[width=.75\columnwidth]{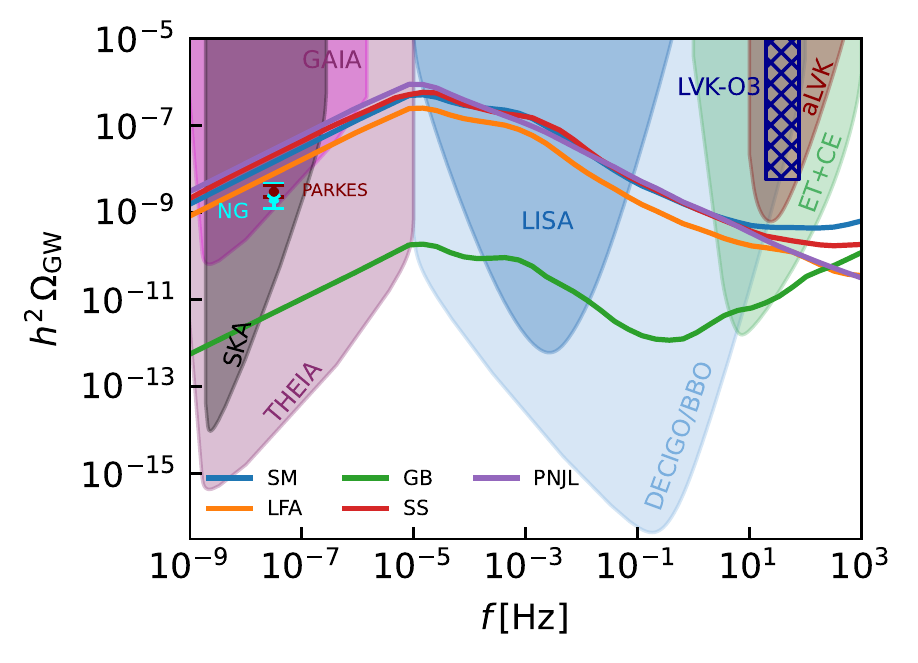}
	\end{center}
	\caption{\label{fig:th} Effects of different thermal history models on the SGWB.}
\end{figure}

We now analyze the imprints of different scenarios of the thermal evolution of the Universe on the SGWB. Our results for the models in Fig.~\ref{fig:fpbh} are shown in Fig.~\ref{fig:th}.

As can be seen, the resulting spectrum of the SGWB is very different depending on the thermal history model considered. In particular, not only does the amplitude change significantly but also the overall spectral shape of $\Omega_{\rm GW}$ is modified. With all the other parameters fixed, these changes are due to the different relative importance of the peaks in $f_{\rm PBH}(M)$. Indeed, the two main peaks affect the normalization $R_{\rm clust}$ of the merger rate and, therefore, the overall amplitude of the SGWB. On the other hand, the relative importance of all the four peaks in $f_{\rm PBH}(M)$ affects its spectral shape.

For the LFA and PNJL cases, the main peak at $M\sim2\,M_\odot$ is less prominent than the other models. Therefore, to be consistent with the merger rate inferred by the LVK collaboration in the range $[5,\,100] M_\odot$, the normalization of the merger rate has to be larger, and they thus have a  larger amplitude at small frequencies. However, they can be distinguished because the PNJL model, for which the features in $f_{\rm PBH}(M)$ for $M\gtrsim5 M_\odot$ are smoothed out, does not show the bump around $f\sim10^{-3}$ Hz. Furthermore, both the PNJL and the LFA model have a spectrum which is significantly different from the SM at the frequencies of ground based detectors, again due to the different structure of the peak around $M\sim100 M_\odot$.  The Solitosynthesis model has a mass function that resembles the SM one the most, and its SGWB is very similar to the one of the SM by red-tilting the primordial power spectrum. Finally, the GB model, which has a mass function with a very large peak at $M\sim1.5-2 M_\odot$, has a much smaller amplitude at low frequencies but shows a faster rise towards high frequencies. However, it cannot be detected by future ground-based detectors.

An important remark is in order. The phase transitions occurring between MeV and GeV, such as those considered here, are responsible for producing a second SGWB. The latter is generically described by a broken power-law peaking around the scales probed by PTA and THEIA, and the location and amplitude of the peak depend on the specific phase transition considered~\cite{Garcia-Bellido:2021zgu}. Therefore, the total SGWB is the superposition of such a broken power law and the one produced by PBH binary mergers presented above. Although it is out of the scope of this work, it would be interesting to study the properties of two backgrounds, such as their spectral shape and duty cycle~\cite{Regimbau:2011bm}, to see how they can be separated. Their joint detection would provide stronger support to our model.

\section{Reconstructing the spectral shape of the SGWB with the PLS approach}
\label{sec:pls}

The results of the previous Section clearly show the relevance of future GW detectors to test our scenario across a very wide range of frequencies. In particular, we have demonstrated how several physical mechanisms leave distinct imprints on the SGWB by distorting its spectral shape in different ways. Therefore, an interesting question that arises quite naturally is whether we can hope or not to reconstruct such peculiar spectral shapes using future observations of the SGWB. Before answering such a question, it is important to mention that sophisticated techniques to reconstruct the frequency dependence of the SGWB have been recently developed within the LISA collaboration~\cite{Caprini:2019egz,Flauger:2020qyi} and in other works~\cite{Karnesis:2019mph,Pieroni:2020rob,Karnesis:2021tsh}. While such techniques are the state of the art of the reconstruction of the SGWB, we adopt in this paper the much simpler (and as such more qualitative) approach of the Power Law Sensitivity curve (PLS) introduced in Ref.~\cite{Thrane:2013oya} and further generalized to the case of SGWB with a non-trivial spectral shape.  This technique is helpful to show qualitatively at which level GW detectors can reconstruct the signal. 

\begin{figure}
	\begin{center}
		\includegraphics[width=\columnwidth]{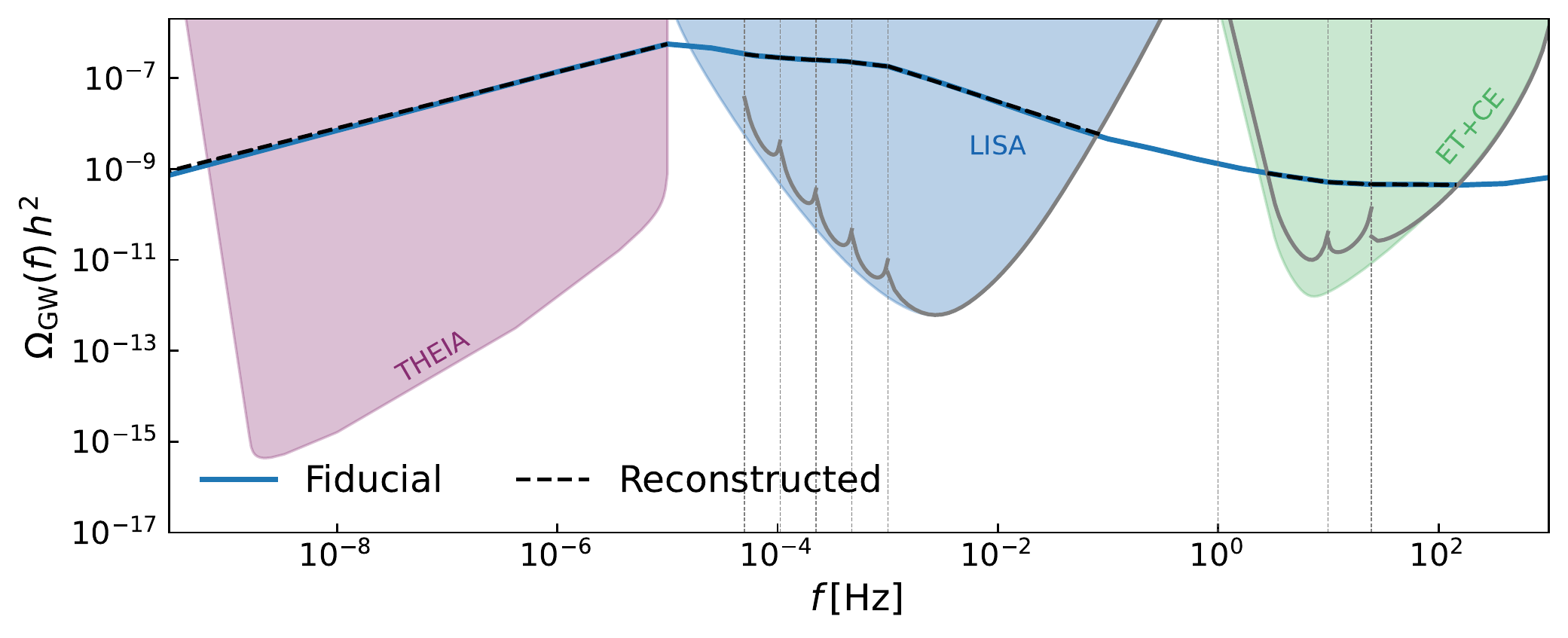}
	\end{center}
	\caption{\label{fig:PLS}  We show the fiducial (blue solid line) and the reconstructed (black dashed line) SGWB in the SM case. Also shown are the PLS curves for THEIA, LISA and ET+CE. The LISA and ET+CE PLS curves are divided into $5$ and $3$ smaller bins respectively, within which the signal can be fitted quite well by a power-law.}
\end{figure}

Our results are shown in Fig.~\ref{fig:PLS} for the case of the SM with $\tilde{n}_s=0.97$, $\tilde{\alpha}_s=\alpha_c=\alpha_z=0$. 
We now describe in details the procedure used to obtain it and how it should be interpreted. 

The meaning of the PLS curve is that every power-law SGWB signal lying above it can be detected and thus reconstructed with a SNR larger than a certain threshold ${\rm SNR}_{\rm thr}$ (we utilize ${\rm SNR}=10$ in this work). For a given frequency range $[f_ {\rm min},\,f_{\rm max}]$ and a given experiment, the curve is constructed using its noise power spectral density and expression of the SNR: 
\begin{equation}
	{\rm SNR}_{\rm thr}=\sqrt{T\,\int_{f_{\rm min}}^{f_{\rm max}}df\,\left(\frac{\Omega_{\rm GW}(f)}{\Omega_s(f)}\right)^2    }
\end{equation}
where $T$ is the observational time and $\Omega_s(f)$ is the energy density calculated from the noise power spectral density. Then for a range of positive and negative values of $\alpha$, one computes the value of the amplitude $A_*$ in $\Omega_{\rm GW}=A_* f^\alpha$ that gives ${\rm SNR}_{\rm thr}$. The PLS curve consists in the envelope of the largest values of $A_*f^\alpha$ at each frequency in the range $[f_ {\rm min},\,f_{\rm max}]$.

It is now easy to understand Fig.~\ref{fig:PLS}, where, for simplicity, we have only picked up THEIA, LISA, and ET+CE among all the future GW detectors. When the signal is above the PLS in a certain frequency range, we fit it to a power-law and draw the fit in dashed black lines. As can be seen, since the signal is very well described by a power-law with spectral index $\alpha_{\rm THEIA}=2/3$ in the frequency range of THEIA, the PLS approach suggests that the signal can be faithfully reconstructed in the frequency range $\sim[10^{-9},\,10^{-5}]$~Hz for THEIA. 

In the LISA and ET+CE bands, however, the signal takes a more complicated form and cannot be described by a simple power law. 
We, therefore, follow Ref.~\cite{Caprini:2019egz} and bin the LISA sensitivity range for the signal to be well approximated by a power-law within each bin. For each bin, then, we construct the PLS. As noticed in Ref.~\cite{Caprini:2019egz}, this results in a degraded sensitivity wrt the PLS computed using the full frequency range. As can be seen, the PLS approach suggests that our signal could be reconstructed quite faithfully also in these ranges of frequencies. An interesting consequence is that if a ground-based experiment ever measures the SGWB of this model, we could tell it apart from PBH scenarios with peaked mass functions. Indeed, the latter gives rise to a SGWB, which typically shows a bump in the frequency range of ground-based detectors, see e.g. Refs.~\cite{Clesse:2016ajp,Raidal:2017mfl,Raidal:2018bbj,Wang:2019kaf,Mukherjee:2021ags,Mukherjee:2021itf}.   

Note that our choice of binning is chosen by eye, simply requiring that the reconstructed signal visually matches the fiducial one. The more sophisticated SGWBinner approach of Ref.~\cite{Caprini:2019egz} iteratively selects the width and number of each bin by using statistical arguments based on information criteria. The results of this Section clearly show the need to apply such advanced reconstruction tools, opening to the possibility to test PBHs and the thermal history of the Universe through the reconstruction of the SGWB.

\section{Conclusions} \label{sec:conclusion}

In this paper, we have discussed in detail the imprints of several physical effects on the Stochastic Gravitational-Wave Background (SGWB) produced by the merger of Primordial Black Hole (PBH) binaries. While it has long been known that such a SGWB can be used to test the clustering properties of PBHs, we have shown that it also encodes precious information about physics occurring at much earlier times. 

In this study, we have assumed a plateau form of the inflationary power spectrum at scales smaller than those probed by the CMB, which we allow to be slightly tilted and distorted by a running parametrization. In this scenario, PBH formation is enhanced at certain mass scales because the critical threshold for density perturbations to collapse to PBHs decreases with the equation of state (EOS) parameter. Since the EOS decreases when particles become massive during the early Universe, several features are imprinted in the mass function of PBHs. Such features are also imprinted in the resulting background, which turns out to cover a very broad range of frequencies.  Therefore, the SGWB signal will have a better chance of being detected by exploiting multi-frequency-band observations of the SGWB, which is one of the main focuses of our paper.

For these reasons, our results are particularly relevant to test Beyond the Standard Model (BSM) scenarios that modify the number of relativistic degrees of freedom in the early Universe and thus the EOS. With the help of a few selected models, we have shown that the resulting SGWB is clearly distinguishable from that produced in the SM and can therefore be used to stringently test such models with upcoming GW observations. In particular, most of the BSM scenarios considered here are also responsible for a SGWB produced by cosmic strings and/or domain walls, which could have a larger amplitude than that from PBHs binaries and is absent in the SM. The simultaneous detection of these two types of SGWB by future GW detectors at different frequencies can therefore be a smoking gun of such scenarios. 

To test the detectability of the GW signal, we have compared the SGWB spectra with the Power-Law Sensitivity (PLS) curves of GW detectors. Our analysis gives us an optimistic take on reconstructing the signal across a very wide range of frequencies, pointing to the exciting possibility of probing the thermal history of the Universe through observations of the SGWB. In order to do so a careful computation of the SGWB taking into account other contributions from the collapse of large overdensities into PBHs~\cite{Clesse:2018ogk}, phase transitions around the QCD epoch \cite{Garcia-Bellido:2021zgu}, and large scalar perturbations at second order and close hyperbolic encounters  of PBHs~\cite{Garcia-Bellido:2021jlq}. We leave this for a future work.

Finally, we would like to mention that our study might also have important consequences for detecting individual merger events by LISA, since the individual GW event can have a quite large amplitude at those scales.

\section*{Acknowledgments}
We thank Eleni Bagui and Sebastien Clesse for very useful discussions and for sharing their draft \cite{Bagui:2021dqi}.  We also thank the IFT Gravitational Wave group for useful comments at various stages of this work and Germano Nardini for comments on the draft. This work is supported by the Spanish Research Project PGC2018-094773-B-C32 (MINECO-FEDER) and the Centro de Excelencia Severo Ochoa Program SEV-2016-0597. MB and SK are supported by the Atracci\'on de Talento contract no. 2019-T1/TIC-13177 granted by the Comunidad de Madrid in Spain and the I+D grant PID2020-118159GA-C42 of the Spanish Ministry of Science and Innovation. SK is partially supported by Japan Society for the Promotion of Science (JSPS) KAKENHI Grant no. 20H01899 and 20H05853. 



\end{document}